\documentclass{elsart}

\usepackage{amsmath,amssymb}

\newcommand\C{\mathbb C}
\newcommand\R{\mathbb R}
\newcommand\diag{\operatorname{diag}}
\newcommand\Det{\operatorname{Det}}

\begin{document}

\begin{frontmatter}

\title{Phylogenetic invariants for stationary base composition }
\thanks[work]{We thank the organizers of the December 2003 ARCC 
Workshop on Computational Algebraic Statistics, and its sponsors, 
AIM and NSF.}

\author[ESA]{Elizabeth S. Allman},
\ead{eallman@maine.edu}
\author[JAR]{John A. Rhodes\thanksref{work}}
\ead{jrhodes@bates.edu}

\address[ESA]{Department of Mathematics and Statistics\\University of 
Southern Maine\\Portland, Maine 04104}
\address[JAR]{Department of Mathematics\\Bates College\\Lewiston, Maine 04240}

\date{June 14, 2004}

\begin{abstract} Changing base composition during the evolution of
biological sequences can mislead some of the phylogenetic inference
techniques in current use.  However, detecting whether such a process
has occurred may be difficult, since convergent evolution may lead to
similar base frequencies emerging from different lineages. 

To study this situation, algebraic models of biological sequence
evolution are introduced in which the base composition is fixed
throughout evolution.  Basic properties of the associated algebraic
varieties are investigated, including the construction of some
phylogenetic invariants.  \end{abstract} 


\end{frontmatter}

\section{Introduction} Models of molecular evolution, such as for DNA
sequences, typically assume evolution occurs along a bifurcating tree,
proceeding from a root representing the common ancestral sequence,
toward the leaves representing the descendent sequences. At each site in
the sequence, bases mutate according to a probabilistic process that
depends upon the edge of the tree. Usually only the sequences at the
leaves of the tree can be observed, while sequences at internal nodes
correspond to hidden variables in this graphical model.  A fundamental
problem of sequence-based phylogenetics is to infer the tree topology
from observed sequences, assuming some reasonable model. 

In the works of Cavendar and Felsenstein \cite{CF87} and Lake
\cite{Lake87}, the connections between this problem and algebraic
geometry first emerged in the phylogenetics literature.  Under many
standard models of molecular evolution, for a fixed tree topology the
joint distribution of bases in the leaf sequences are described by
polynomial equations in the parameters of the model, thus parameterizing
a variety associated to the tree. The defining polynomials of this
variety, called \emph{phylogenetic invariants}, are polynomials that
vanish on any joint distribution arising from the tree and model,
regardless of parameter values.  Finding phylogenetic invariants for
various models has been of interest both for providing theoretical
understanding, and in hopes that methods of phylogenetic inference that
do not require parameter estimation may be developed. See \cite{Fel03}. 

For certain models, much progress has been made on determining
invariants. Key advances for group-based models such as the Kimura
three-parameter, were made in \cite{MR93m:62121} and \cite{SteSzErWa93},
which built on the Hadamard conjugation introduced in \cite{HenPen93}.
Recently, Sturmfels and Sullivant \cite{q-bio.PE/0402015} further
exploited the Hadamard conjugation to recognize these varieties were
toric, completing the determination of all invariants in this case. For
the general Markov model, Allman and Rhodes \cite{AR03}, found new
constructions of invariants, though the complete determination of the
ideal is still open. 

In this paper, we consider models that lie between group-based models
and the general Markov model.  Specifically, we assume that a fixed
vector describes the relative frequencies of the bases in sequences at
\emph{every} node of the tree, so that the base composition of sequences
remains stable throughout evolution. 

Our motivation for this assumption is a biological one.  Many of the
models currently assumed in performing inference with real data make an
assumption of a stable base composition (e.g., all group-based models,
the general time-reversible model). However, there are data sets in
which base composition seems to have changed during evolution, as
reflected in comparisons of the sequences at the leaves. Although the
extent to which this issue is problematic in real data sets is
controversial, a number of authors have pointed out that changing base
composition may mislead some methods of inference, especially if it
results in convergent mutations in different parts of the tree.  See
\cite{LSHP,ConLew,RosKum} and their references. 

In \cite{KumGad} a `disparity index' was introduced as a simple
statistical test that might indicate inhomogeneity of the mutation
process along the different edges of the tree.  This index is based on a
pair-wise comparison of base compositions of sequences at the leaves.
It is, however, possible that all leaf sequences have the same base
composition, while an internal node sequence has a different one. 
Indeed, this is
exactly the issue with convergent mutations; base composition may appear
to be the same in observed sequences, yet it differed in the common
ancestral sequence. If a model is chosen only through comparing the base
compositions of sequences at the leaves, it may be an inappropriate one. 

Better understanding the constraints placed on the joint distribution of
bases in sequences from various taxa by an assumption of a stable base
distribution is therefore desirable.  To begin investigating this issue,
in section \ref{sec:models} we introduce three models of molecular
evolution that include an assumption of stable base distribution. When
the number of bases in the model is $\kappa=2$, the three models are the
same, and its structure allows us to give a more in-depth analysis than
for general $\kappa$. This is the subject of section \ref{sec:k2}, where
we give a rational map inverting the parameterization, and find the full
ideal of phylogenetic invariants for the 3-taxon tree. In section
\ref{sec:karb}, the case of general $\kappa$ is considered. Basic facts
about the associated phylogenetic varieties, such as their dimension and
irreducibility/non-irreducibility, are investigated. Although our
knowledge of phylogenetic invariants is incomplete, we give
constructions of some for these models.

\section{The Models}\label{sec:models}

Let $T$ denote an undirected bifurcating tree, with $n$ leaves labeled
by the taxa $a_1, a_2,\dots, a_n$. If $r$ is some vertex in $T$, either
internal or terminal, we use $T_r$ to denote the tree rooted at $r$. We
view $T_r$ as a directed graph, with all edges directed away from $r$
forming a set $Edge(T_r)$. Thus $T_r$ represents a hypothetical
evolutionary history of the taxa in their descent from a common ancestor
at $r$. For simplicity, we refer to $T_r$ as a \emph{rooted $n$-taxon
tree}. 

We model the evolution along $T_r$ of sequences composed from an
alphabet $[\kappa]=\{1,2,\dots, \kappa \}$ of bases or states (e.g.,
$\kappa=4$ for DNA). A \emph{root distribution vector} $\mathbf
p_r=(p_1,p_2,\dots, p_\kappa)$, with $ p_i\in[0,1]$, $\sum_i p_i=1$,
describes the frequency of bases in an ancestral sequence.  To each $e
\in Edge(T_r)$ we associate a $\kappa\times \kappa$ Markov matrix $M_e$
(with entries in $[0,1]$, each row summing to 1) whose $(i,j)$-entry
specifies the conditional probability of base $i$ at the initial vertex
mutating to base $j$ at the final vertex of the edge. Together $\mathbf
p_r$ and $\{M_e \}_{e\in Edge(T_r)}$ comprise the parameters of the
model.  If no additional requirements are placed on $\mathbf p_r$ or the
$M_e$, then we have described the general Markov model (GM) of sequence
evolution, studied in \cite{AR03}. 

Letting $X_{GM,\kappa,T_r}$ denote the parameter space for the
$\kappa$-state GM model on $T_r$, we can view $X_{GM,\kappa,T_r}$ as a
subset of $[0,1]^M$ for $M=\kappa+\kappa^2E$ with $E=2n-3$, the number
of edges of $T$. We have a map $$\phi=\phi_{GM,\kappa,
T_r}:X_{GM,\kappa,T_r}\to [0,1]^{\kappa^n}\subset \C^{\kappa^n},$$ so
that $\phi(x)$ gives the joint distribution of bases in aligned
sequences at the leaves arising from the parameter choice $x$.
Specifically, $\phi(x)=P=(p_{j_1j_2\dots j_n})$, a $\kappa\times\dots
\times \kappa$ tensor with entries $$p_{j_1j_2\dots j_n}= \sum_{i\in
\mathcal I(j_{1},j_{2},\dots, j_{n})} p_{i_r} \prod_{\substack{e\in
Edge(T_r),\\ e=(v\to w)}} M_e(i_v,i_w), $$ where $\mathcal
I(j_{1},j_{2},\dots, j_{n})=\left\{ (i_v) ~|~ v\in Vert(T_r), i_v\in
[\kappa], i_{a_k}= j_{k} \right \}\subset [\kappa]^{2n-2}.  $ Note
$\phi$ is a polynomial map, viewed as a function of the entries of
$\mathbf p_r$ and $M_e$, and extends to a polynomial map $\C^M\to
\C^{\kappa^n}$ which we also denote by $\phi$.

In this paper we are interested in submodels of the GM model, in which
we have stable base frequencies at all vertices in the tree.  We
introduce three such models, defined by imposing additional restrictions
on parameters of the GM model. After formally defining the models, we will
motivate their assumptions and names.

\begin{itemize}

\item Stable Base Distribution Model (SBD): 1) $\mathbf p_r$ has no zero
entries, and 2) $\mathbf p_r$ is fixed by all $M_e$; that is, $\mathbf
p_r M_e =\mathbf p_r$ for all edges $e$.

\item Simultaneous Diagonalization Model (SD): In addition to the
assumptions of SBD, 3) with $D_r=\diag (\mathbf p_r)$, all matrices in
$$\{M_e ~|~ e \in Edge(T_r) \} \cup \{ D_r^{-1} M_e^T D_r ~|~ e \in
Edge(T_r) \}$$ commute with one
another. 

\item Algebraic Time Reversible Model (ATR): In addition to the
assumptions of SD, 4) for all edges $e$, $M_e=D_r^{-1} M_e^T D_r$. 

\end{itemize}

We will also need:

\begin{defn}For any model $\mathcal M$ formed from the $\kappa$-state GM
model by imposing additional assumptions on the parameters, and for any
rooted $n$-taxon tree $T_r$, we let $X_{\mathcal M,\kappa, T_r}$ denote
the parameter space of $\mathcal M$ on $T_r$.  Then the algebraic
variety $V(\mathcal M,\kappa,T_r)$ is the Zariski closure in
$\C^{\kappa^n}$ of $\phi(X_{\mathcal M,\kappa, T_r})$.  \end{defn}

We now expand upon the model definitions.  First, assuming $\mathbf p_r$
has no zero entries, the matrices $D_r^{-1} M_e^T D_r$ are also Markov
matrices fixing $\mathbf p_r$. They arise naturally as follows: Consider
a 2-taxon tree consisting of a single edge $e$ from vertex $r$ to vertex
$s$, with model parameters $\mathbf p_r$ and $M_e$. Then the joint
distribution of bases in aligned sequences at $r$ and $s$ arising from
these parameter choices is given by the entries in $D_rM_e$. Assuming
$M_e$ fixes $\mathbf p_r$, so that $\mathbf p_r$ is also the base
distribution of a sequence at $s$, then the identity $$D_r M_e = (D_r
(D_r^{-1} M_e^T D_r))^T$$ shows that the model parameters $\mathbf p_r,
M_e$ on the 1-edge tree rooted at $r$ lead to the same joint
distribution as the model parameters $\mathbf p_r, D_r^{-1} M_e^T D_r$
on the 1-edge tree rooted at $s$.  More generally, as shown in
\cite{SSH94,AR03} for the GM model, parameters for the SBD, SD, or ATR
model on $T_r$ produce the same joint distribution as the set of
parameters on $T_s$ for any other vertex $s$, simply by defining
$\mathbf p_s=\mathbf p_r$, and for those edges whose directions have
reversed in changing the root location, replacing $M_e$ by $ D_r^{-1}
M_e^T D_r$. In particular, we see

\begin{prop} $\phi(X_{\mathcal M,\kappa, T_r})$ and $V(\mathcal M,
\kappa, T_r)$ are independent of the choice of $r$ for $\mathcal M$=SBD,
SD, and ATR. Thus, for these models, $V(\mathcal M, \kappa, T)$ is
well-defined without reference to $r$.  \end{prop}

Second, the requirement for the SD and ATR model that the specified
collection of matrices commute is, in fact, equivalent to an assumption
that those matrices are simultaneously diagonalizable. To see this,
first note that the commutation assumption is equivalent to the
commutation of the collection $$\{D_r^{1/2} M_e D_r^{-1/2} ~|~ e \in
Edge(T_r) \} \cup \{ D_r^{-1/2} M_e^T D_r^{1/2} ~|~ e \in Edge(T_r)
\}.$$ But this implies in particular that each matrix $D_r^{1/2} M_e
D_r^{-1/2}$ is normal, and hence diagonalizable.  Commutativity then
implies the existence of simultaneous eigenvectors for this collection,
and hence for the original collection. Conversely, if the matrices are
simultaneously diagonalizable, they certainly commute.

Third, the ATR model is related to the general time-reversible model
(GTR) often used in phylogenetic studies.  The GTR assumes that for each
edge $e$, $M_e=\exp(Rt_e)$, where $t_e$ is a scalar parameter and $R$ is a
rate matrix (with rows summing to 0) common to all edges with the
properties that $D_r R$ is symmetric and $\mathbf p_r R=\mathbf 0$
(\cite{Fel03}). A collection of Markov matrices arising from GTR
parameters thus satisfies the hypotheses of the ATR model. However, the
common rate matrix assumption of the GTR imposes a relationship among
the logarithms of the eigenvalues of the Markov matrices $M_e$ which the
ATR does not, and thus the ATR is more amenable to algebraic analysis. 

Finally, we note that the group-based models, such as the Kimura
3-parameter (KST), can be viewed as the ATR together with additional
assumptions on the eigenvectors of the $M_e$. For instance, KST requires
the eigenvectors be the columns of a $4\times4$ Hadamard matrix, with
$(1,1,1,1)$ the stable base distribution. 

We summarize the relationships of the various algebraic models with the
inclusions $$\text{Group-based} \subset \text{ATR} \subseteq \text{SD}
\subseteq \text{SBD} \subset \text{GM}.$$ For $\kappa\ge 3$, these
inclusions are all strict, though for $\kappa=2$ the three central
models are identical, as shown below. Of course, the associated varieties
are related by a reversed chain of inclusions. 

\section{The 2-base model}\label{sec:k2}

For $\kappa=2$, the SBD, SD, and ATR models are all the same.  To see
this, and fix notation for future use, consider the SBD model, with root
distribution vector $\mathbf p_r=(p,1-p)=(p,q)$. Since each matrix $M_e$
has left eigenvector $\mathbf p_r$ and right eigenvector $(1,1)$, both
with eigenvalue 1, we readily find we can express
$$M_{e}=M(m_e)=\begin{pmatrix} 1-m_eq & m_e q\\ m_ep & 1-m_ep
\end{pmatrix}, $$ thus associating a single scalar parameter $m_e$ to
each edge.  We also see that $M_e$ satisfies the hypotheses of the ATR
model as well. (In fact, for $\kappa=2$, the ATR model and the GTR model
also coincide.) The form of $M_e$ allows us to identify parameters for
an $n$-taxon tree with a point $(p;\{m_e\})\in \R^{2n-2}$.

We first consider a 3-taxon tree $T_r$, rooted at its central node, with
three edges $e_1$, $e_2$, and $e_3$ leading from $r$ to leaves $a_1$,
$a_2$, and $a_3$.  Labeling the states 0 and 1, and using these as
indices to refer to matrix entries corresponding to the states, the
joint distribution of bases at a site in sequences at the leaves is now
described by a $2\times2\times2$ tensor $P=(p_{ijk})$, where $$p_{ijk}=
pM_1(0,i)M_2(0,j)M_3(0,k) + q M_1(1,i)M_2(1,j)M_3(1,k).  $$

Viewing $P$ as a polynomial function of $p,m_1,m_2,m_3$, we thus have a
map $\varphi:\C^4 \to \C^8$, and readily see that the Zariski closure of
the image of $\varphi$ is $V(SBD,\kappa,T)$. 

For notational ease, we follow the convention that replacing an index by
the symbol `+' indicates marginalization over that index. For instance,
$p_{ij+}=\sum_k p_{ijk}$, while $p_{i++}=\sum_{j,k} p_{ijk}$. 

\begin{prop} \label{prop:invert} The following rational map provides an
explicit inverse to the parameterization of the map $\varphi$:
\begin{align*}
p &= p_{0++}, \\
m_i &=1-\frac{\sum_{ijk} (-1)^{i+j+k}p_{ijk} p_{(1-i)++} p_{(1-j)++} p_{(1-k)++}}
{ (p_{1++}-p_{0++}) d_i},\ \ i=1,2,3,
\end{align*}
where $d_1=\det(p_{+ij})$, $d_2=\det(p_{i+j})$, and $d_3=\det(p_{ij+}).$

\end{prop}

\begin{pf} Define a $2\times2\times 2$ diagonal tensor $D$ with
$D(0,0,0)=p$, $D(1,1,1)=q$, and all other entries zero. We then have
\begin{equation} p_{ijk}=\sum_{l,m,n=0}^1
D(l,m,n)M_1(l,i)M_2(m,j)M_3(n,k),\label{eq:Pparam} \end{equation}
expressing $P$ as the result of an action of an element of $GL_2\times
GL_2\times GL_2$ on $D$.  Also observe that each matrix $M_i$ has as
right eigenvectors $(1,1)$ and $(-q,p)$, with eigenvalues $1$ and
$1-m_i$, respectively.  Thus multiplying the tensor $P$, whose entries
are polynomials in $p$, $m_1$, $m_2$, and $m_3$, by the vector $\mathbf v
=(v_0,v_1)=(-q,p)$ along each of its indices, yields \begin{align}
g_0&=\sum_{i,j,k=0}^1 p_{ijk} v_i v_j v_k \label{eq:g0}\\ &=
\sum_{i,j,k=0}^1 \sum_{l,m,n=0}^1 D(l,m,n) M_1(l,i) v_i M_2(m,j) v_j
M_3(n,k) v_k.\notag \end{align}

Interchanging summations, and using that $\mathbf v$ is an eigenvector
of each of the $M_i$ yields
\begin{align}
g_0
&= \sum_{l,m,n=0}^1 (1-m_1)(1-m_2)(1-m_3)D(l,m,n) v_l v_m v_n\notag\\
&=(1-m_1)(1-m_2)(1-m_3)p q(p-q)\label{eq:g0param}.
\end{align}

Multiplying similarly, with two copies of $\mathbf v$ and one of $(1,1)$
yields
\begin{align}
g_1=\sum_{i,j,k=0}^1 p_{ijk} v_jv_k=(1-m_2)(1-m_3)p q,\notag\\
g_2=\sum_{i,j,k=0}^1 p_{ijk} v_iv_k=(1-m_1)(1-m_3)p q,\label{eq:gi}\\
g_3=\sum_{i,j,k=0}^1 p_{ijk} v_iv_j=(1-m_1)(1-m_2)p q.\notag
\end{align}

Now since $p=p_{0++}$, if we express the entries of $\mathbf v$ as
linear polynomials in the $p_{ijk}$, we may view the $g_i$ as
polynomials in the $p_{ijk}$ as well. Then from equations (\ref{eq:g0})
and (\ref{eq:gi}) we see that $g_0$ is of degree $4$, while $g_1$,
$g_2$, and $g_3$ are each of degree 3. 

A calculation shows all four of these polynomials have a factor of
$s=\sum_{i,j,k=0}^1 p_{ijk}$, which of course evaluates to 1 on $V$, so
we may replace each $g_i$ with $\tilde g_i=g_i/s$, if desired.  We also
note that explicit expressions for the quadratic $\tilde g_i$ as
ordinary matrix determinants can be given: $$\tilde g_1 = \det
(p_{+ij}),\ \tilde g_2 = \det (p_{i+j}) , \ \tilde g_3 = \det (p_{ij+}
).$$

Equations (\ref{eq:g0param}) and (\ref{eq:gi}) now lead directly to
formulas for the $m_i$, $$ m_i=1-\frac {g_0}{(v_0+v_1)g_i},\text{ for
$i=1,2,3$,}$$ which yield the stated map. \qed 

\end{pf}

The explicit invertibility of the parameterization map for the 3-taxon
tree readily extends to $n$-taxon trees. 

\begin{thm} Suppose $T_r$ is a rooted $n$-taxon tree with
$(p;\{m_e\}_{e\in Edge(T_r)})\in \C^{2n-2}$ defining $\mathbf
p_r=(p,1-p)$, $M_e=M(m_e)$ and $$P=(p_{i_1i_2\dots i_n})=\phi(\mathbf
p_r; \{M_e\}_{e\in Edge(T_r)} ).$$ Then the polynomial map
$\varphi:(p;\{m_e\}_{e\in Edge(T_r)})\mapsto P$ is inverted by a
rational map explicitly given by the formulas:

1) $p=p_{0++\dots+}$.

2) For each terminal edge $e_0$, assume without loss of generality that
$e_0=(v\to a_1)$. Choose two other taxa $a_2,a_3$ such that the path
from $a_2$ to $a_3$ in $T$ passes through $v$. Then
$$m_{e_0}=1-\frac{\sum_{ijk}(-1)^{i+j+k} p_{ijk+\dots+}
p_{(1-i)+\dots+} p_{(1-j)+\dots+} p_{(1-k)+\dots+}} {
(p_{1+\dots+}-p_{0+\dots+}) \det(p_{+ij+\dots+})}.$$

3) For each internal edge $e_0=(v\to w)$, chose four taxa which, without
loss of generality, we assume are $a_1,a_2,a_3,a_4$, such that the path
joining $a_1$ to $a_2$ in $T$ passes through $v$, but not through $w$;
and the path joining $a_3$ to $a_4$ passes through $w$, but not $v$.
Then $$m_{e_0}=1- \frac{ (\sum_{ijk}(-1)^{i+j+k} p_{ijk+\dots+}
p_{(1-i)+\dots+} p_{(1-j)+\dots+}
p_{(1-k)+\dots+})\det(p_{+i+j+\dots+})}{ (\sum_{ijk}(-1)^{i+j+k}
p_{+ijk+\dots+} p_{(1-i)+\dots+} p_{(1-j)+\dots+}
p_{(1-k)+\dots+})\det(p_{ij+\dots+})}.$$

\end{thm}

\begin{pf} The formula for $p$ is clear.  For the remaining formulas,
note that because we are dealing with the $ATR$ model, the root location
may be changed without changing the variety, and while moving the root
may change a direction of an edge $e$, the matrix $M_e$ is unchanged. 

For a terminal edge $e_0$ as described above, the 3-dimensional tensor
$(p_{ijk+\dots+})=\phi(\mathbf p;M_1,M_2,M_3)$ for a 3-taxon tree
$T'_v$, where $M_i=\prod_{e\in Path(v,a_i)} M_e$ with $Path(v,a_i)$ the
set of edges in the path joining $v$ to $a_i$. In particular,
$M_1=M_{e_0}$, so applying the formula of Proposition \ref{prop:invert}
for $m_1$ yields the desired formula.

Similarly, for an internal edge $e_0$ as described above, start with the
3-dimensional tensor $(p_{ijk+\dots+})=\phi_{T'}(\mathbf p;M_1,M_2,M_3)$
for the 3-taxon tree $T'_v$. Then, since
$M(m)M(m')=M(m'')$ is equivalent to $(1-m)(1-m')=1-m''$, by applying the
formula of Proposition \ref{prop:invert}, we find $$\prod_{e\in
Path(v,a_3)} (1-m_{e}) =\frac{\sum_{ijk} (-1)^{i+j+k} p_{ijk+\dots+}
p_{(1-i)+\dots+} p_{(1-j)+\dots+}
p_{(1-k)+\dots+}}{(p_{1+\dots+}-p_{0+\dots+}) \det(p_{ij+\dots+})}.$$

Likewise, considering the 3-dimensional tensor $(p_{+ijk+\dots+})$, we
find $$\prod_{e\in Path(w,a_3)} (1-m_e)=\frac{\sum_{ijk}
(-1)^{i+j+k}p_{+ijk+\dots+} p_{(1-i)+\dots+} p_{(1-j)+\dots+}
p_{(1-k)+\dots+}}{(p_{1+\dots+}-p_{0+\dots+}) \det(p_{+i+k+\dots+})}.$$

Since $\prod_{e\in Path(v,a_3)} (1-m_e)=(1-m_{e_0}) \prod_{e\in
Path(w,a_3)} (1-m_e)$, this yields the given formula.\qed 

\end{pf}

Now, to determine phylogenetic invariants for the SBD model with $\kappa =2$,
we first consider the 3-taxon tree $T$.  We seek all polynomials in the
$p_{ijk}$ that vanish on $\varphi(\C^4)$, and thus define $V(SBD,2,T)$. 

As we are considering a submodel of GM, we obtain
the stochastic invariant, which defines $V(GM,2,T)$: $$f_0=1- p_{+++}.$$

Several other invariants for the stable base composition model are
easily found.  The distribution of bases in a sequence at $a_i$ is given
by the vector $\mathbf p_i$ where $$\mathbf p_1=(p_{i++}),\ \mathbf
p_2=(p_{+i+}),\ \mathbf p_3=(p_{++i}).$$ Since each leaf sequence must
have the same base composition for SBD, we set $\mathbf p_1-\mathbf
p_2=\mathbf p_1- \mathbf p_3=\mathbf 0,$ obtaining two linear invariants
$$f_1=p_{010}+p_{011} - p_{100}-p_{101},\ \ f_2=p_{001}+p_{011}
-p_{100}-p_{110},$$ whose span includes that arising from $\mathbf
p_2-\mathbf p_3=0$.  These are the invariants underlying the disparity
index of \cite{KumGad}. 

From equations (\ref{eq:g0param}) and (\ref{eq:gi}) we can also see that
\begin{equation} h=\tilde g_0^2 v_0 v_1 +\tilde g_1 \tilde g_2 \tilde
g_3 (v_0+v_1)^2 \label{eq:hinv} \end{equation} is an invariant of degree
8. However, while $h\notin(f_0,f_1,f_2)$, the ideal $(f_0,f_1,f_2,h)$ is
not the full ideal of invariants.  Using Macaulay 2 \cite{M2} to find
the kernel of the ring map associated to $\varphi$ quickly yields a single
invariant of degree 6 with 258 terms, which together with $f_0$, $f_1$,
and $f_2$ generates the full ideal defining $V(SBD,2,T)$. 

In fact, this invariant can be explained through the hyperdeterminants
of \cite{MR95e:14045}. For a $2\times 2 \times 2$ tensor such as $P$,
the hyperdeterminant is given explicitly as \begin{multline*} 
\Det(P)=(p_{000}^2p_{111}^2+p_{001}^2p_{110}^2+p_{010}^2p_{101}^2+p_{011}^2p_{100}^2)
\\ -2(p_{000}p_{001}p_{110}p_{111}+
p_{000}p_{010}p_{101}p_{111}+p_{000}p_{011}p_{100}p_{111} \\
+p_{001}p_{010}p_{101}p_{110}+p_{001}p_{011}p_{110}p_{100}+p_{010}p_{011}p_{101}p_{100})\\
+4(p_{000}p_{011}p_{101}p_{110}+p_{001}p_{010}p_{100}p_{111}).
\end{multline*}

Now reasoning from equation (\ref{eq:Pparam}) and using the invariance
properties of $\Det(P)$ under the $SL_2\times SL_2\times SL_2$ action,
one finds that in terms of model parameters, $$\Det(P)=
p^2q^2(1-m_1)^2(1-m_2)^2(1-m_3)^2.$$ Thus $\tilde g_1 \tilde g_2 \tilde
g_3 -pq\Det(P)=0,$ and so $\tilde g_1 \tilde g_2 \tilde g_3 +v_1
v_0\Det(P),$ viewed as a degree 6 polynomial in the $p_{ijk}$, is an
invariant.  Expressing this explicitly in terms of the $p_{ijk}$, we
have the invariant $$f_3= \det(p_{+ij}) \det(p_{i+j}) \det(p_{ij+})
                 - p_{0++}  p_{1++}  \Det(p_{ijk}).$$

A computation with Macaulay 2 now yields the following:

\begin{thm} \label{thm:SBD2id} The ideal of phylogenetic invariants
vanishing on $V(SBD,2,T)$ for the 3-taxon tree is $$(f_0,f_1,f_2,f_3).$$
\end{thm}

We thank a reviewer for pointing out that the $2\times2\times2$
hyperdeterminant was introduced into a phylogenetic context in
\cite{SJ04}, where it is called the \emph{tangle}. That paper considers
the 2-base GM model on a 3-taxon tree rooted along an edge, and proposes
the hyperdeterminant as a generalized `distance.'

For an $n$-taxon tree, determining the full ideal of invariants for the
2-base SBD model remains open.  Of course, this model inherits the
invariants of the GM model, which have been conjectured in \cite{PS} to
be generated by `edge invariants' arising from rank conditions on
2-dimensional flattenings of the tensor. This issue for GM will be dealt
with in \cite{ARgm}. Additional invariants for SBD arise from applying
the invariants of Theorem \ref{thm:SBD2id} to all 3-dimensional
marginalizations of the $n$-dimensional tensor $P$.  One might suspect
these generate the full ideal, but even for the $4$-taxon tree we have
been unable to confirm this computationally. 

\section{The $\kappa$-base models, arbitrary $\kappa$}\label{sec:karb}

\begin{prop}
Let $T$ be an  $n$-taxon tree, with $E=E(n)=2n-3$ the number of its edges.
Denoting the dimension of the variety
$V(\mathcal M, \kappa, T)$ by $d(\mathcal M, \kappa, T)$,
\begin{gather*}
d(SBD,\kappa,T)= (\kappa-1) +(\kappa-1)^2 E,\\
d(SD,\kappa,T) =d(ATR,\kappa,T)= \frac{\kappa(\kappa-1)}2 +(\kappa-1)E. 
\end{gather*}
\end{prop}

\begin{pf} For fixed $\kappa$ and $T$, choose a root $r$ for $T$.  For
each model $\mathcal M$ we consider here, the parameter space
$X_{\mathcal M,\kappa,T_r}\subset X_{GM,\kappa, T_r}$ is a semialgebraic
subset of $\R^M$ with $M=\kappa+\kappa^2 E$, as all our model
assumptions are polynomial equalities or inequalities placing
restrictions on the entries of $p_r$ and the $M_e$. 

For each $(\mathcal M, \kappa, T_r)$, we will find a complex
quasi-projective variety $X$ of dimension
$d=d(\mathcal M, \kappa, T)$ and a generically finite map $\psi:X\to
\C^M$, such that $X_{\mathcal M,\kappa,T_r}\subset \psi(X)$ and
$\overline{ \psi^{-1}(X_{\mathcal M,\kappa,T_r})}= X$.  These conditions
imply $$\overline {\psi(X)} = \overline{X_{\mathcal M,\kappa,T_r}},$$ so
applying the map $\phi=\phi_{GM,\kappa,T_r}:\C^M\to \C^{\kappa^n}$, we
find $$\overline{\phi\circ \psi(X)}=\overline{\phi(X_{\mathcal
M,\kappa,T_r})}=V(\mathcal M,\kappa,T).$$ Since the results of
\cite{AR03} show $\phi$ is generically finite, the general fiber
$(\phi\circ\psi)^{-1}(P)$ is of dimension zero.  Using a standard result
on the dimension of fibers of regular maps (see \cite{Harris}, for
instance),  we conclude that the dimension of $X$ and $V(\mathcal M,
\kappa, T)$ are the same. 

We begin with the SBD model, so $d=(\kappa-1) +(\kappa-1)^2 E$.  Let
$$X=\{\mathbf x\in \C^d ~|~ \sum_{i=1}^{\kappa-1} x_i \ne 1\},$$ and
define the map $\psi$ as follows: $\psi(x)=(\mathbf p,\{M_e\})$ where
$p_i=x_i$ for $i=1,\dots,\kappa -1$, and the upper left
$(\kappa-1)\times (\kappa-1)$ blocks of each $M_e$ are given by
successive entries in $\mathbf x$.  Use the conditions that $\sum_i
p_i=1$, $\sum_j M_e(i,j)=1$, and $\mathbf p M_e =\mathbf p$ to give
rational formulas for the remaining entries of $\mathbf p$ and $M_e$ in
terms of $\mathbf x$.  Clearly $\psi$ is 1-1 and $X_{SBD,\kappa,T_r}
\subset \psi(X)$.  Moreover, $\psi^{-1}(X_{SBD,\kappa,T_r})$ is dense in
$X$ since it contains a Euclidean-open subset of the real points of $X$,
which is Zariski dense in $X$. 

\smallskip

For the ATR model, with $d=\frac{\kappa(\kappa-1)}2+(\kappa-1)E$, let
$$X=\{(Q,\mathbf u)~|~ Q\in O_\kappa(\C),\ Q=(q_{ij}),\ q_{i1}\ne 0,\
\mathbf u \in \C^{(\kappa-1)E} \}.$$ Here $O_\kappa(\C)$ is the
variety of complex orthogonal $\kappa\times\kappa$ matrices, which has
dimension $\frac {\kappa(\kappa-1)}2$. Define $\psi$ by: $\psi(Q,\mathbf
u)=(\mathbf p,\{M_e\})$ where $\mathbf p=(q_{11}^2,q_{21}^2,\dots,
q_{\kappa 1}^2)$, and, with $D=\diag(q_{11},q_{21}\dots, q_{\kappa 1})$,
$$M_{e_i}=D^{-1} Q\diag(1,x_{j+1},x_{j+2},\dots,x_{j+\kappa-1})Q^T D,$$
where $j=(\kappa-1)(i-1)$.  That $\psi$ is generically finite is clear,
and that $X_{ATR,\kappa, T_r} \subseteq \psi(X)$ follows from the
discussion in section \ref{sec:models}. Also, the set
$\psi^{-1}(X_{ATR,\kappa, T_r})$ is dense in $X$ since it contains a
Euclidean-open subset of the real points of $X$, which is Zariski dense
in $X$. 

\smallskip

Finally, for the SD model, recall \cite{JacII} that a family of real
commuting normal matrices $A_{e_i}$ can be simultaneously expressed as
$A_i=QB_iQ^T$, with $Q\in O_\kappa(\R)$, and the $B_i$ real block diagonal
matrices with the same block structure, where each diagonal block is
either $1\times 1$ or $2\times 2$ of the form $\left ( \begin{smallmatrix}
a&b\\-b&a\end{smallmatrix} \right )$. The block structures we need to consider will
have $n$ $1\times 1$ blocks, the first of which is 1, followed by $m$
$2\times 2$ blocks, where $n\ge 1$, $m\ge 0$, and $n+2m=\kappa$.
Proceeding similarly to the case of the ATR model, for each of these
$\lfloor \frac{\kappa+1}{2} \rfloor$ possible block structures $\mathcal
B$, we let $X_{\mathcal B}$ denote a copy of $X$ as defined for ATR, and
define a map $\psi_{\mathcal B}:X_{\mathcal B}\to \C^M$ similar to the
ATR map, where the entries in $\mathbf u$ give the independent block
entries in $B_i$ and $M_{e_i}= D^{-1} Q B_i Q^T D.$ Letting $X$ be the
disjoint union of the $X_{\mathcal B}$, we obtain a map $\psi:X\to
\C^M$.  The rest of the argument is similar to that for the ATR model.
\qed \end{pf}

The construction of the varieties $X$ and maps $\psi$ in this proof also
yield

\begin{prop} The varieties $V(SBD,\kappa,T)$ and $V(ATR,\kappa,T)$ are
irreducible, but $V(SD,\kappa,T)$ is the union of $\lfloor \frac
{\kappa+1}2 \rfloor$ distinct irreducible components, one of which is
$V(ATR,\kappa,T)$.  \end{prop}

For each of the SBD, SD, and ATR $\kappa$-base models on a tree $T$, we
can construct a few phylogenetic invariants, though we are far from a
full understanding of the ideals and varieties.  Since any submodel
inherits all invariants of a supermodel, and $SBD\supseteq SD\supseteq
ATR$, we consider the models in that order. In addition, since these are
all submodels of GM, all GM invariants on $T$, such as those of
\cite{AR03,ARgm}, are also invariants of these models.

\noindent {\bf SBD model:} We first consider the 3-taxon tree $T_r$,
rooted at the central node and reason similarly to section \ref{sec:k2}.
If $(p_{ijk})= \phi(\mathbf p_r;M_1,M_2,M_3)$, then we have
$$p_{i++}=p_{+i+}=p_{++i},$$
giving $2(\kappa-1)$ independent linear invariants expressing equality
of base distributions at the leaves.  We can also construct an invariant
from the hyperdeterminant $\Det(p_{ijk})$ on $\kappa \times \kappa\times
\kappa$ tensors. Letting $m=m(\kappa)$ denote the degree of this
polynomial (so $m(3)=36$ and $m(4)= 272$), then as before we find
$$\Det(p_{ijk}) = (\det(M_1) \det(M_2) \det(M_3)\det(D_r) )^{m/\kappa}.$$
Similarly,
\begin{gather*}
\det(p_{ij+})= \det(M_1)\det(M_2)\det(D_r), \\ 
\det(p_{i+j})= \det(M_1)\det(M_3)\det(D_r), \\ 
\det(p_{+ij})= \det(M_2)\det(M_3)\det(D_r),
\end{gather*}
so
$$ (\det(p_{ij+})\det(p_{i+j})\det(p_{+ij})) ^{m/(2\kappa)}-
(\prod_i p_{i++}) ^{m/(2\kappa)}   \Det(p_{ijk})$$
is an invariant for the SBD model, since $2\kappa$ divides $m$,
as can be shown from formulas in \cite{MR95e:14045}.

To see this is not an invariant for the GM model, we check that it does
not vanish for some GM parameters.  Indeed, if the parameters are chosen
so the entries of $\mathbf p_r$ and $p_{i++}$ are positive, the $M_i$
are non-singular, and $\det(D_r)\ne \prod_i p_{i++}$, then the invariant
will be non-zero. 

A similar construction replacing $\Det$ with any relative invariant $h$
of $GL_\kappa \times GL_\kappa \times GL_\kappa$ acting on $\mathbb
C^{\kappa} \otimes \mathbb C^{\kappa} \otimes \mathbb C^{\kappa}$
produces a phylogenetic invariant, provided $h$ does not vanish on all
diagonal tensors. If $h$ does vanish on diagonal tensors, then $h$ is
already an invariant of the GM model. 

To obtain $n$-taxon invariants we can of course compose 3-taxon
invariants with any marginalization map of $n$-dimensional tensors to
3-dimensional ones. 

\smallskip

\noindent {\bf SD model:} Note that for any choice of 2 taxa $a_j,a_k$
on $T$, if $P=\phi(\mathbf p_r,\{M_e\})$, where $(\mathbf
p_r,\{M_e\})\in X_{SD,\kappa,T_r},$ then the 2-dimensional
marginalization $\widetilde P^{jk}$ of $P$ obtained by summing over
indices corresponding to all other taxa will be of the form $D_r M$,
where $M$ is a product of matrices in the collection $$\{M_e ~|~ e \in
Edge(T_r) \} \cup \{ D_r^{-1} M_e^T D_r ~|~ e \in Edge(T_r) \},$$ and
$D_r=\diag(\mathbf p_r)=\diag(p_{1+\dots+}, p_{2+\dots+},\dots,
p_{\kappa+\dots+}).$ Thus all matrices in the collection
$$\{D_r^{-1}\widetilde P^{jk} ~|~ 1 \le j < k \le n \} \cup \{ D_r^{-1}
(\widetilde P^{jk})^T ~|~ 1 \le j < k \le n \}$$ will commute.  For each
pair chosen from this set, we get a collection of polynomials of degree
$\kappa+1$ from the statement of commutativity: For instance,
$$(D_r)^{-1} \widetilde P^{jk} (D_r)^{-1} \widetilde P^{lm}= (D_r)^{-1}
\widetilde P^{lm} (D_r)^{-1} \widetilde P^{jk},$$ gives invariants from
the entries of $$\widetilde P^{jk} (\det(D_r) (D_r)^{-1}) \widetilde
P^{lm}- \widetilde P^{lm} (\det(D_r) (D_r)^{-1}) \widetilde P^{jk}.$$
That some of these are not invariants of the SBD model when $\kappa>2$
can be verified, most easily for a 2-taxon tree by a generic choice of
SBD parameters. 

\smallskip

\noindent {\bf ATR model:} We consider first a 2-taxon tree, with $P\in
\phi(X_{ATR,\kappa,T_r})$. Then $P=D_r M_e$, so the condition
$M_e=D_r^{-1}M_e^T D_r$ implies $P=P^T$. The entries of this matrix
equation then give linear invariants, which are not invariants of the SD
model for $\kappa>2$, since there exist parameters for the SD model with
$M_e \ne D_r^{-1}M_e^T D_r$. Composing these invariants with
2-dimensional marginalization maps gives linear invariants for an
$n$-taxon tree.

\bibliographystyle{alpha}
\bibliography{phylo}

\end{document}